\documentclass{iopart}
\usepackage{bbm,amssymb,amsbsy,graphicx,subfigure,times,float}
\usepackage{hyperref}
\usepackage[latin1]{inputenc}
\usepackage{color}

\newcommand{\R}{\mathbbm{R}}

\newcommand{\id}{\mathbbm{1}}

\renewcommand{\tr}{{\rm Tr}\,}
\renewcommand{\det}{{\rm Det}\,}
\newcommand{\gr}[1]{\boldsymbol{#1}}
\newcommand{\be}{\begin{equation}}
\newcommand{\ee}{\end{equation}}
\newcommand{\bea}{\begin{eqnarray}}
\newcommand{\eea}{\end{eqnarray}}

\newcommand{\sig}{\gr{\sigma}}

\newcommand{\eq}[1]{Eq.~(\ref{#1})}

\begin{document}
%%%%%%%%%%%%%%%%%%%%%%%%%%%%%%%%%%%%%%%%%%%%%%%%%%%%%%%%%%%
\title[Standard forms and entanglement of Gaussian states under local operations]
{Standard forms and entanglement engineering of multimode Gaussian states under local operations}
\author{Alessio Serafini}
\address{Institute for Mathematical Sciences, Imperial College London, 53 Prince's Gate,
SW7 2PE, United Kingdom and QOLS, Blackett Laboratory,
Imperial College London, Prince Consort Road, SW7 2BW, United
Kingdom}
\author{Gerardo Adesso}
\address{Dipartimento di Fisica ``E. R.
Caianiello'', Universit\`a degli Studi di Salerno; CNR-Coherentia,
Gruppo di Salerno; and INFN Sezione di Napoli-Gruppo Collegato di
Salerno; Via S. Allende, 84081 Baronissi (SA), Italy}

\begin{abstract}
We investigate the action of local unitary operations on multimode
(pure or mixed) Gaussian states and single out the minimal number of
locally invariant parametres which completely characterise the
covariance matrix of such states. For pure Gaussian states, central
resources for continuous-variable quantum information, we
investigate separately the parametre reduction due to the additional
constraint of global purity, and the one following by the
local-unitary freedom. Counting arguments and insights from the
phase-space Schmidt decomposition and in general from the framework
of symplectic analysis, accompany our description of the standard
form of pure $n$-mode Gaussian states. In particular we clarify why
only in pure states with $n \le 3$ modes all the direct correlations
between position and momentum operators can be set to zero by local
unitary operations. For any $n$, the emerging minimal set of
parametres contains complete information about all forms of
entanglement in the corresponding states.  An efficient
 state engineering scheme (able to encode direct
correlations between position and momentum operators as well) is
proposed to produce entangled multimode Gaussian resources, its
number of optical elements matching the minimal number of locally
invariant degrees of freedom of general pure $n$-mode Gaussian
states. Finally, we demonstrate that so-called ``block-diagonal'' Gaussian
states, without direct correlations between position and momentum,
are systematically less entangled, on average, than arbitrary pure
Gaussian states.
\end{abstract}


%%%%%%%%%%%%%%%%%%%%%%%%%%%%%%%%%%%%%%%%%%%%%%%%%%%%%%%%%%%
\maketitle
%%%%%%%%%%%%%%%%%%%%%%%%%%%%%%%%%%%%%%%%%%%%%%%%%%%%%%%%%%%
\section{Prologue}

Entanglement between the subsystems of composite quantum systems is
arguably one of the most radical features of quantum mechanics, the
one invoking a dramatic departure from classical principles
\cite{schr}. This is probably one of the reasons why a fully
satisfactory understanding and characterisation of such a feature in
the most general setting is still lacking. Accordingly, the task of
developing a comprehensive theoretical framework to qualify and
quantify multipartite entanglement stands as a major issue to be
achieved in quantum physics. A novel insight into the role of
entanglement in the description of quantum systems has been gained
through the quantum information perspective, mostly focusing on the
usefulness of entanglement, rather than on its mathematical
characterization. In these years, quantum entanglement has turned
from a paradoxical concept into a physical resource allowing for the
encoding, manipulation, processing and distribution of information
in ways forbidden by the laws of classical physics. In this respect,
entanglement between canonically conjugate continuous variables (CV)
of infinite-dimensional systems, like harmonic oscillators, light
modes and atomic ensembles, has emerged as a versatile and powerful
resource \cite{eisplenio}. In particular, multimode Gaussian states
have been proven useful for a wide range of implementations in CV
quantum information processing \cite{brareview}, and advances in the
characterisation of their bipartite and multipartite entanglement
have recently been recorded \cite{adebook}. In experiments, one
typically aims at preparing pure states, with the highest possible
entanglement, even though unavoidable losses and thermal noises will
affect the purity of the engineered resources, and hence the
efficiency of the realised protocols \cite{francesi}. It is
therefore important to understand the structure of correlations in
pure Gaussian states, and to provide `economical' schemes to produce
such states in the lab with minimal elements, thus reducing the
possibility of accumulating errors and unwanted noise.

Gaussian states of CV systems are special in that they are
completely specified by the first and second moments of the
canonical bosonic operators. However, this already reduced set of
parametres (compared to a true infinite-dimensional one needed to
specify an arbitrary non-Gaussian CV state) contains many redundant
degrees of freedom which have no effect on the entanglement. A basic
property of multipartite entanglement is in fact its invariance
under unitary operations performed locally on the subsystems. To
describe entanglement efficiently, is thus natural to lighten
quantum systems of the unnecessary degrees of freedom adjustable by
local unitaries (LUs), and to classify states according to {\em
standard forms} representative of LU equivalence classes
\cite{linden}. When applied to Gaussian states of $n$ modes, the freedom arising
from the LU invariance immediately rules out the vector of first
moments, which can be arbitrarily adjusted by local displacements in
phase space (LUs on the Hilbert spaces) and thus made null without
any loss of generality. One is then left with the $2n(2n+1)/2$ real
parametres constituting the symmetric covariance matrix (CM) of the
second moments (rigorously defined in the following).

In this paper, we study the action of LU operations on a general CM
of a multimode Gaussian state. We compute the minimal number of
parametres which completely characterise Gaussian states, up to LUs.
The set of such parametres will contain {\em complete} information
about any form of bipartite or multipartite entanglement in the
corresponding Gaussian states. We give accordingly the standard form
of the CM of a (generally mixed) $n$-mode Gaussian state. We then
focus on pure states, the preferred resources for CV quantum
communication and information processing, and study how the
additional constraint of global purity leads to a further reduction
of the minimal set of LU invariant parametres. We interpret those
degrees of freedom in terms of correlations between the canonical
operators of the various modes, and discuss how to
engineer pure $n$-mode Gaussian states starting from a two-mode
squeezed state and $n-2$ single-mode squeezed beams, via passive
operations only. Our results generalise the classification of Ref.
\cite{generic}, where the standard form of $n$-mode pure Gaussian
states with no correlations between position ($\hat x$) and momentum
($\hat p$) operators was given, together with an optimal scheme to
engineer such ``block-diagonal'' resources (employed in most CV
quantum information protocols) in an optical setting. In this
respect, we show that nonzero $\hat x$-$\hat p$ correlations lead to
an enhancement of the typical entanglement in the sense of
\cite{typical}.

%%%%%%%%%%%%%%%%%%%%%%%%%%%%%%%%%%%%%%%%%%%%%%%%%%%%%%%%%%%
\section{Technical introduction}\label{techno}

We consider systems described by
pairs of canonically conjugated operators $\{\hat{x}_{j},\hat{p}_{j}\}$
with continuous spectra,
acting on a tensor product of infinite dimensional Hilbert spaces.
Let $\hat R = (\hat x_{1},\hat
p_{1},\ldots,\hat x_{n},\hat p_{n})$ denote the vector of the
operators $\hat x_{j}$ and $\hat p_{j}$. The canonical commutation
relations for the $\hat R_{i}$ can be expressed in terms of the
symplectic form ${\Omega}$
\[
[\hat R_{j},\hat R_k]=2i\Omega_{jk} \; ,
\]
\[
{\rm with}\quad{\Omega}\equiv \bigoplus_{j=1}^{n}
{\omega}\; , \quad {\omega}\equiv \left( \begin{array}{cc}
0&1\\
-1&0
\end{array}\right) \; .
\]

The state of a CV system can be equivalently described by a
positive trace-class operator (the density matrix $\varrho$) or by
quasi--probability distributions.
Throughout the paper, we shall focus on
states with Gaussian characteristic functions and
quasi-probability distributions, commonly referred to as `Gaussian
states'.
By definition, a Gaussian state $\varrho$ is completely
characterised by the first and second statistical moments of the
canonical operators. We will just consider states with null first moments, completely
determined by the symmetric
covariance matrix (CM) $\sig$ with entries
$
\sig_{jk}\equiv \tr \left[\varrho (\hat{X}_j \hat{X}_k +
\hat{X}_k \hat{X}_j)\right]
$.
Being the variances and covariances of quantum operators,
such entries are obtained by noise variance and noise correlation
measurements (obtained by `homodyne' detection for optical systems).
They can be expressed as energies by multiplying
them by the quantity $\hbar \omega$, where $\omega$ is the
frequency of the considered mode. In fact, for any $n$-mode state
the quantity $\hbar \omega\tr({\sig}/4)$ is just the
contribution of the second moments to the average of the ``free''
Hamiltonian $\sum_{i=1}^{n} (a^{\dag}_ia_i + 1/2)$.

Let us recall some useful results about symplectic operations, along with their consequences on the
description of Gaussian states.
Being positive definite \cite{serafozzijosab}, the CM of a $n$--mode Gaussian state can
always be written as
\begin{equation}
\sig=S^T \nu S \; , \label{willia}
\end{equation}
with $S\in Sp_{(2n,\mathbb{R})}$ and
\begin{equation}
\nu=\,{\rm diag}({\nu}_{1},{\nu}_{1},\ldots,{\nu}_{n},{\nu}_{n}) \, ,
\label{therma}
\end{equation}
corresponding to the CM of a tensor product of states at thermal
equilibrium with local temperatures $T_{j}=2(\nu_{j}-1)$. The
quantities $\{\nu_{j}\}$ are referred to as the {\em symplectic
eigenvalues} of the CM $\sig$, the transformation $S$ is said to
perform a {\em symplectic diagonalisation} of $\sig$, while the
diagonal matrix with identity blocks $\nu$ is referred to as the
{\em Williamson form} of $\sig$ \cite{willy}. The symplectic
eigenvalues $\{\nu_{j}\}$ can be determined as the positive square
roots of the eigenvalues of the positive matrix
$-\Omega\sig\Omega\sig$. Such eigenvalues are in fact invariant
under the action
of symplectic transformations on the matrix $\sig$.\\
We briefly remark that all the entropic quantities of Gaussian states can be expressed in
terms of their symplectic eigenvalues. Notably, the
`purity' $\tr{\varrho^2}$ of a Gaussian state $\varrho$ is simply
given by the symplectic invariant
$\det{\sig}=\prod_{i=1}^{n}\nu_i$, being $\tr{\varrho^2} =
(\det{\sig})^{-1/2}$.

Central to our analysis will also be the following general decomposition
of a symplectic transformation $S$ (referred to as the ``Euler'' or ``Bloch-Messiah'' decomposition
\cite{pramana,braunsqueezirreducibile}):
\begin{equation}
S = O' Z O ,\label{euler1}
\end{equation}
where $O, O' \in K(n)= Sp(2n,\R)\cap SO(2n)$ are orthogonal
symplectic transformations, while
$$
Z=\oplus_{j=1}^{n}\left(\begin{array}{cc}
z_j & 0 \\
0 & \frac{1}{z_j}
\end{array}\right) \; ,
$$
with $z_{j}\ge 1$ $\forall$ $j$. The set of such $Z$'s forms a
non-compact subgroup of $Sp_{2n,\R}$ comprised of local
(single-mode) squeezing operations (borrowing the terminology of
quantum optics, where such transformations arise in degenerate
parametric down-conversion processes). Moreover, let us also mention
that the compact subgroup $K(n)$ is isomorphic to the unitary group
$U(n)$, and is therefore characterised by $n^2$ independent
parametres. To acquaint the reader with the flavour of the counting
arguments which will accompany us through this paper (and with the
nontrivial aspects contained therein), let us combine the Williamson
and the Euler decomposition to determine the number of degrees of
freedom of an arbitrary {\em mixed} $n$-mode Gaussian state (up to first moments),
thus obtaining $n+2n^2+n-n=2n^2+n$. The first two addenda are just the sum
of the number of symplectic eigenvalues ($n$) and of degrees of freedom of a
symplectic operation ($2n^2+n$, resulting from two symplectic orthogonal transformations and
from $n$ single-mode squeezing parametres).
Finally, the subtracted $n$ takes into account the invariance under
single-mode rotations of the local Williamson forms (which `absorbs'
one degree of freedom {\em per mode} of the symplectic operation
describing the state according to \eq{willia}). Actually, the
previous result is just the number of degrees of freedom of a
$2n\times2n$ symmetric matrix (in fact, the only constraint $\sig$
has to fulfill to represent a physical state is the semidefinite
$\sig+i\Omega \ge 0$, which compactly expresses the uncertainty
relation for many modes \cite{seraprl}).

Finally, we recall the form of the CM $\sig^{2m}$ of a {\it two-mode
squeezed state}:

\be
\sig^{2m}=\left(\begin{array}{cccc}
\cosh r&0&\sinh r&0\\
0&\cosh r&0&-\sinh r\\
\sinh r&0&\cosh r&0\\
0&-\sinh r&0&\cosh r
\end{array}\right) \; , \label{2msq}
\ee parametrised by the positive squeezing $r$.  This class of
states represents the prototype of CV entanglement both for the
experimentalist (it can be generated by non-degenerate ``parametric
down conversion'') and for the theorist (it encompasses, in the
limit $r\rightarrow \infty$, the perfectly correlated seminal
Einstein-Podolsky-Rosen state \cite{EPR}) and will play a crucial
role in several arguments to follow.

%%%%%%%%%%%%%%%%%%%%%%%%%%%%%%%%%%%%%%%%%%%%%%%%%%%%%%%%%%%
\section{Standard forms of mixed states}\label{sform}

Before addressing the reductions of {\em pure} states, let us briefly consider
the standard forms of general {\em mixed} $n$-mode Gaussian states under local, single-mode symplectic operations.
Let us express the CM $\sig$ in terms of $2\times2$ sub-matrices $\sig_{jk}$,
defined by
$$
\sig \equiv \left(\begin{array}{ccc}
\sig_{11} &\cdots &\sig_{1n}\\
\vdots & \ddots &  \vdots \\
\sig_{1n}^{\sf T} & \cdots  & \sig_{nn}
\end{array}\right) \; \
$$
each sub-matrix describing either the local CM of mode $j$ ($\sig_{jj}$)
or the correlations between the pair of modes $j$ and $k$ ($\sig_{jk}$).

Let us remind the reader of
the Euler decomposition of a generic single-mode symplectic transformation $S_1({\vartheta',\vartheta'',z})$:
$$
S_{1}({\vartheta',\vartheta'',z}) = \left(\begin{array}{cc}
\cos{\vartheta'} & \sin{\vartheta'} \\
-\sin{\vartheta'} & \cos{\vartheta'}
\end{array}\right)
\left(\begin{array}{cc}
z & 0 \\
0 & \frac1z
\end{array}\right)
\left(\begin{array}{cc}
\cos{\vartheta''} & \sin{\vartheta''} \\
-\sin{\vartheta''} & \cos{\vartheta''}
\end{array}\right) \,
$$
into two single-mode rotations (``phase shifters'', with reference to the
``optical phase'' in phase space) and one squeezing operation.
We will consider the reduction of a generic CM $\sig$ under local operations
of the form $S_{l} \equiv \bigoplus_{j=1}^{n} S_{1}(\vartheta'_{j},\vartheta''_j,z_j)$.
The local symmetric blocks $\sig_{jj}$ can all be diagonalised by the first rotations
and then symplectically diagonalised
({\em i.e.}, made proportional to the identity) by the subsequent squeezings,
such that $\sig_{jj}=a_j \id_{2}$ (thus reducing the number of parametres in each
diagonal block to the local symplectic eigenvalue, determining the
entropy of the mode).
The second series of local rotations can then be applied to manipulate
the non-local blocks, while leaving the local ones unaffected (as
they are proportional to the identity).
Different sets of $n$ entries in the non-diagonal sub-matrices can be thus
set to zero. For an even total number of
modes, all the non-diagonal blocks $\sig_{12}$, $\sig_{34}$,\ldots,$\sig_{(n-1)n}$
describing the correlations between disjoint pairs of quadratures can be
diagonalised (leading to the singular-value diagonal form of each block), with no conditions on all the other blocks.
For an odd number of modes, after the diagonalisation of the blocks relating disjoint
quadratures,
a further non-diagonal block involving the last mode (say, $\sig_{1n}$)
can be put in triangular form by a rotation on the last mode.

Notice finally that the locally invariant degrees of freedom of a generic Gaussian state of $n$
modes are $(2n+1)n-3n=2n^2-2n$,
as follows from the subtraction of
the number of free parametres of the local symplectics from the one of a generic state
-- with an obvious exception for $n=1$, for which the number of free
parametres is $1$, due to the rotational invariance of single-mode Williamson
forms (see the discussion about the vacuum state in Sec.~\ref{redu}).

%%%%%%%%%%%%%%%%%%%%%%%%%%%%%%%%%%%%%%%%%%%%%%%%%%%%%%%%%%%

\section{Degrees of freedom of pure Gaussian states}\label{condpuri}

Pure Gaussian states are characterised by CMs with Williamson form
equal to the identity. As we have seen, the Williamson decomposition
provides a mapping from any Gaussian state into the uncorrelated
product of thermal (generally mixed) states: such states are pure
(corresponding to the vacuum), if and only if all the symplectic
eigenvalues are equal to $1$.

The symplectic eigenvalues of a generic CM $\sig$
are determined as the eigenvalues of the matrix $|i\Omega\sig|$,
where $\Omega$ stands for the symplectic form.
Therefore, a Gaussian state of $n$-modes with CM $\sig$ is pure if and only if
\be
-\sig\Omega\sig\Omega = \id_{2n} \; . \label{puri}
\ee

It will be convenient here to reorder the CM, and to decompose it in
the three sub-matrices $\sig_x$, $\sig_p$ and $\sig_{xp}$, whose
entries are defined as \be \hspace*{-1.5cm} (\sig_x)_{jk} =
\tr[\varrho \hat{x}_j \hat{x}_k]\;,\quad (\sig_p)_{jk} = \tr[\varrho
\hat{p}_j \hat{p}_k]\;,\quad (\sig_{xp})_{jk} = \tr[\varrho
\{\hat{x}_j,\hat{p}_k\}/2]\;, \ee such that the complete CM $\sig$
is given in block form by \be \sig = \left(\begin{array}{cc}
\sig_x & \sig_{xp} \\
\sig_{xp}^{\sf T} & \sig_{p}
\end{array}\right) \; . \label{sig}
\ee Let us notice that the matrices $\sig_{x}$ and $\sig_p$ are
always symmetric and strictly positive, while the matrix $\sig_{xp}$
does not obey any general constraint.

Eqs.~(\ref{puri}) and (\ref{sig}) straightforwardly lead to the
following set of conditions \bea
\sig_{x}\sig_{p} = \id_n + \sig_{xp}^2 \; , \label{first} \\
\sig_{xp}\sig_{x} - \sig_{x}\sig_{xp}^{\sf T} = 0 \; , \label{second} \\
\sig_{p}\sig_{x} = \id_n + \sig_{xp}^{\sf T\,2} \; , \label{redu1} \\
\sig^{\sf T}_{xp}\sig_{p} - \sig_{p}\sig_{xp} = 0 \; . \label{redu2}
\eea
Now, \eq{redu1} is obviously obtained by
transposition of \eq{first}. Moreover, from (\ref{first}) one
gets \be \sig_{p} = \sig_{x}^{-1}(\id_{n}+\sig_{xp}^2) \; ,
\label{sigp} \ee while \eq{second} is equivalent to \be
\sig_{x}^{-1}\sig_{xp} - \sig_{xp}^{\sf T}\sig_{x}^{-1} = 0
\label{secondbis} \ee (the latter equations hold generally, as
$\sig_{x}$ is strictly positive and thus invertible).
Eq.~(\ref{secondbis}) allows one to show that any $\sig_{p}$
determined by (\ref{sigp}) satisfies the condition (\ref{redu2}).
Therefore, only Eqs.~(\ref{first}) and (\ref{second}) constitute
independent constraints and fully characterise the CM of pure
Gaussian states.

Given any (strictly positive) matrix $\sig_{x}$ and (general) matrix
$\sig_{xp}$, the fulfillment of condition (\ref{second}) allows to
specify the second moments of any pure Gaussian state, whose
sub-matrix $\sig_p$ is determined by \eq{sigp} and does not involve
any additional degree of freedom.

A straightforward counting argument  thus yields the number of
degrees of freedom of an arbitrary pure Gaussian state\footnote{The
same number could have been inferred, via an essentially identical
reasoning, from the normal form of pure Gaussian states
independently derived in Ref.~\cite{gaussianeof}, Lemma 1.}, by
adding the entries of a general and of a symmetric $n\times n$
matrices and subtracting the equations of the antisymmetric
condition (\ref{second}): $n^2+n(n+1)/2-n(n-1)/2=n^2+n$, in
compliance with the number dictated by the Euler decomposition of a
symplectic operation: \be \sig = S^{\sf T} \id_{2n} S = O^{\sf T}
Z^2 O \; . \ee Notice that, if either $\sig_{x}$ or $\sig_{xp}$ are
kept fixed, the constraint (\ref{second}) is just a linear
constraint on the entries of the other matrix, which can always be
solved (it cannot be overdetermined, since the number of equations
$n(n-1)/2$ is always smaller than the number of variables, either
$n^2$ or $n(n+1)/2$).

A preliminary insight into the role of local operations in
determining the number of degrees of freedom of pure CMs is gained
by analysing the counting of free parametres in the continuous
variable analogue of the Schmidt decomposition. The CM of any pure
$(m+n)$-mode Gaussian state is equivalent, up to local symplectic
transformations on the $m$-mode and $n$-mode subsystems, to the
tensor product of $m$ decoupled two-mode squeezed states (assuming,
without loss of generality, $m<n$) and $n-m$ uncorrelated vacua
\cite{botero}. Besides the $m$ two-mode squeezing parametres, the
degrees of freedom of the local symplectic transformations to be
added are $2n^2+n+2m^2+m$. However, a mere addition of these two
values leads to an overestimation with respect to the number of free
parametres of pure CMs determined above. This is due to the
invariance of the CM in `Schmidt form' under specific classes of
local operations. Firstly, the $(n-m)$-mode vacuum (with CM equal to
the identity) is trivially invariant under local orthogonal
symplectics, which account for $(n-m)^2$ parametres. Furthermore,
one parametre is lost for each two-mode squeezed block with CM
$\sig^{2m}$ given by \eq{2msq}: this is due to an invariance under
single-mode rotations peculiar to two-mode squeezed states. For such
states, the sub-matrices $\sig^{2m}_{x}$ and $\sig^{2m}_{p}$ have
identical -- and all equal -- diagonal entries, while the sub-matrix
$\sig^{2m}_{xp}$ is null. Local rotations embody two degrees of
freedom -- two local `angles' in phase space -- in terms of
operations. Now, because they act locally on $2\times2$ identities,
rotations on both single modes cannot affect the diagonals of
$\sig^{2m}_{x}$ and $\sig^{2m}_{p}$, nor the diagonal of
$\sig^{2m}_{xp}$, which is still equal to zero. In principle, they could thus
lead to two (possibly different) non-diagonal elements for
$\sig^{2m}_{xp}$ and/or to two different non-diagonal elements for
$\sig^{2m}_{x}$ and $\sig^{2m}_{p}$ (which, at the onset, have
opposite non-diagonal elements, see \eq{2msq}), resulting in
$$
\sig^{2m}_{x}=\left(\begin{array}{cc}
a & c_1 \\
c_1 & a
\end{array}\right) \, ,
\sig^{2m}_{p}=\left(\begin{array}{cc}
a & c_2 \\
c_2 & a
\end{array}\right) \, ,
\sig^{2m}_{xp}=\left(\begin{array}{cc}
0 & y \\
z & 0
\end{array}\right)
$$
However, elementary considerations, easily worked out
for such $2\times2$ matrices, show that Eqs.~(\ref{second}) and (\ref{sigp}) imply
$$
c_1=-c_2 \;\; ,\;\; y=z \quad {\rm and} \quad a^2-c_1^2 = 1+y^2 \, .
$$
These constraints reduce from five to two the number of free parametres in
the state: {\em the action of local single-mode rotations --
generally embodying two independent parametres -- on two-mode
squeezed states, allows for only one further independent degree of
freedom}. In other words, all the Gaussian states resulting from
the manipulation of two-mode squeezed states by local rotations
(``phase-shifters'', in the experimental terminology) can be
obtained by acting on only one of the two modes. One of the two
degrees of freedom is thus lost and the counting argument displayed
above has to be recast as $m+2n^2+n+2m^2+m-(m-n)^2-m=(m+n)^2+(m+n)$,
in compliance with what we had previously established.

As we are about to see, this invariance, peculiar to two-mode squeezed states,
also accounts for the reduction of locally invariant
free parametres occurring in pure two-mode Gaussian states.

%%%%%%%%%%%%%%%%%%%%%%%%%%%%%%%%%%%%%%%%%%%%%%%%%%%%%%%%%%%
\section{Reduction under single-mode operations}\label{redu}

Let us now determine the reduction of degrees of freedom achievable
for pure Gaussian states by applying local single-mode symplectic
transformations. Notice that all the entanglement properties (both
bipartite and multipartite) of the states will solely depend on the
remaining parametres, which cannot be canceled out by LU operations.

In general, for $n$-mode systems,
local symplectic operations have $3n$ degrees of freedom, while $n$-mode
pure Gaussian states are specified, as we just saw, by $n^2+n$ quantities.
The subtraction of these two values yields a residual number of parametres
equal to $n^2-2n$.
However, this number holds for $n\ge3$, but fails for single- and two-mode
states. Let us analyse the reasons of this occurrence.

For single-mode systems, the situation is trivial, as one is allowing for
all the possible operations capable, when acting on the vacuum, to unitarily yield any possible state. The number of free parametres is then clearly zero
(as any state can be reduced into the vacuum state, with CM equal to the
$2\times2$ identity). The expression derived above would instead give $-1$. The reason
of this mismatch is just to be sought in the invariance of the vacuum under
local rotations: only two of the three parametres entering the Euler
decomposition actually affect the state. On the other hand, one can also
notice that these two latter parametres, characterising the squeezing and
subsequent last rotation of the Euler decomposition acting on the vacuum,
are apt to completely reproduce any possible single-mode state.
Clearly, this situation is the same as for any $n$-mode pure Gaussian state
under global operations: the first rotation of the Euler decomposition is always irrelevant,
thus implying a corresponding reduction of the free parametres of the state with respect to the
most general symplectic operation.

As for two-mode states, the counting argument above would give zero locally
invariant parametres. On the other hand, the existence of a class of states
with a continuously varying parametre determining the amount of bipartite
entanglement [the two-mode squeezed states of \eq{2msq}],
clearly shows that the number of free parametres cannot be
zero. Actually, local symplectic operations allows one to bring {\em any}
(pure or mixed) two-mode Gaussian state in a ``standard form'' with
$\sig_{xp}=0$ and with identical diagonals for $\sig_{x}$ and $\sig_{p}$.
Imposing then \eq{second} on such matrices, one finds that the only pure
states of such a form have to be two-mode squeezed states.
Therefore, we know that the correct number of locally invariant free parametres has to be one. Even though local symplectic operations on two-mode states
are determined by $6$ parametres, they can only cancel $5$ of the $6$ parametres
of pure two-mode states. This is, again, due to the particular
transformation properties of two-mode squeezed states under single-mode rotations,
already pointed out in Section \ref{condpuri} when addressing the counting
of degrees of freedom in the Schmidt-like decomposition:
local rotations acting on a two-mode squeezed state add only one independent parametre.
The most general two-mode pure Gaussian state results from a two-mode squeezed
state by a single local rotation on any of the two modes, followed by two
local squeezings and two further rotations acting on different modes.
Notice that the same issue arises for $(m+n)$-mode states to be reduced under
local $m$- and $n$-mode symplectic operations. A mere counting of degrees of freedom
would give a residual number of local free parametres equal to $(m+n)^2+m+n-2m^2-2n^2-m-n=
-(m-n)^2$. This result is obviously wrong, again due to a loss of parametres in the transformations
of particular invariant states. We have already inspected this very case and pointed out such invariances
in our treatment of the Schmidt decomposition (previous Section): we know that the number of locally irreducible
free parametres is just $\min(m,n)$ in this case, corresponding to the tensor product
of two-mode squeezed states and uncorrelated vacua.

For $n\ge3$, local single-mode operations can fully reduce the number
of degrees of freedom of pure Gaussian states by their total number of parametres.
The issue encountered for two-mode states does not occur here, as the first
single-mode rotations can act on different non-diagonal blocks of the CM ({\em
i.e.}, pertaining to the correlations between different pairs of modes).
The number of such blocks is clearly equal to $(n^2-n)/2$ while the number
of local rotations is just $n$. Only for $n=1,2$ is the latter value larger
than the former: this is, ultimately, why the simple subtraction of degrees
of freedom only holds for $n\ge3$.
To better clarify this point, let us consider a CM $\sig^{3m}$ in the limiting instance $n=3$.
The general standard form for (mixed) three-mode states implies the conditions (see Sec.~\ref{sform})
\be
{\rm diag}\,(\sig^{3m}_{x}) = {\rm diag}\,(\sig^{3m}_{p})  \label{diagga}
\ee
and
\be
\sig^{3m}_{xp} = \left(\begin{array}{ccc}
0&0&0 \\
0&0&u \\
s&t&0
\end{array}
\right) \; . \label{sparse} \ee The diagonal of $\sig^{3m}_{x}$
coincides with that of $\sig^{3m}_{p}$ (which always results from
the local single-mode Williamson reductions) while six entries of
$\sig^{3m}_{xp}$ can be set to zero.
For pure states, imposing
\eq{second} results into a linear system of three equations for the
nonzero entries of $\sig^{3m}_{xp}$, with coefficients given by the
entries of $\sig^{3m}_{x}$.
Exploiting the complete positivity of $\sig^{3m}_{x}$,
one can show that such a system implies $s=t=u=0$.
Therefore, for pure
three-mode Gaussian states, the matrix $\sig^{3m}_{xp}$ can be set to zero
by local symplectic operations alone on the individual modes.
The entries of the symmetric positive definite matrix
$\sig^{3m}_{x}$ are constrained by the necessity of
Eqs.~(\ref{first}) -- which just determines $\sig^{3m}_{p}$ -- and
(\ref{diagga}), which is comprised of three independent conditions
and further reduces the degrees of freedom of the state to the
predicted value of three. An alternative proof of this is presented
in Ref. \cite{3mpra}.

Let us also incidentally remark that the possibility of reducing the
sub-matrix $\sig_{xp}$ to zero by local single-mode operations is
exclusive to two-mode (pure and mixed) and to three-mode pure
states. This is because, for general Gaussian states, the number of
parametres of $\sig_{xp}$ after the local Williamson
diagonalisations is given by $n(n-1)$ (two per pair of modes) and
only $n$ of these can be canceled out by the final local rotations,
so that only for $n<3$ can local operations render $\sig_{xp}$ null.
For pure states and $n>2$ then, further $n(n-1)/2$ constraints on
$\sig_{xp}$ ensue from the antisymmetric condition (\ref{second}):
this number turns out to match the number of free parametres in
$\sig_{xp}$ for $n=3$, but it is no longer enough to make
$\sig_{xp}$ null for pure states with $n\ge4$.

Summing up, we have rigorously determined the number of  ``locally
irreducible'' free parametres of pure Gaussian states, unambiguously
showing that the quantification and qualification of the
entanglement (which, by definition, is preserved under LU
operations) in such states of $n$ modes is completely determined by
$1$ parametre for $n=2$ and $(n^2-2n)$ parametres for $n>2$.

%\subsection{States with zero $\hat x$-$\hat p$ correlations}

\section{Efficient state engineering of multimode pure Gaussian states}
It would be desirable to associate the mathematically clear number
$(n^2-2n)$ with an operational, physical insight. In other words, it
would be useful for experimentalists (working, for instance, in
quantum optics) to be provided with a recipe to create pure $n$-mode
Gaussian states with completely general entanglement properties in
an `economical' way, in the precise, specific sense that exactly $(n^2-2n)$ optical elements are used.
A transparent approach to develop such a procedure consists in considering the
reverse of the phase space $1 \times (n-1)$ Schmidt decomposition,
as introduced in Section \ref{condpuri}. Namely, a completely
general (not accounting for the local invariances) state engineering
prescription for pure Gaussian states can be cast in two main steps:
(i) create a two-mode squeezed state of modes 1 and 2, which
corresponds to the multimode state in its Schmidt form; (ii) operate
with the most general $(n-1)$-mode symplectic transformation
$S^{-1}$ on the block of modes $\{2,3,\ldots,n\}$ (with modes
$i=3,\ldots,n$ initially in the vacuum state) to redistribute
entanglement among all modes. The operation $S^{-1}$ is the inverse
of the transformation $S$ which brings the reduced CM of modes
$\{2,3,\ldots,n\}$ in its Williamson diagonal form. It is also known
that any such symplectic transformation $S^{-1}$ (unitary on the
Hilbert space) can be decomposed in a network of optical elements
\cite{reckzeil}.\footnote{Notice that, even though Ref.~\cite{reckzeil}
refers to compact (`{\em passive}') transformations alone, the Euler decomposition,
which involves only passive operations and single-mode squeezings, allows one to straightforwardly extend
such decompositions in terms of single- and two-mode operations
to general symplectic transformations.}
The number of elements required to accomplish this
network, however, will in general greatly exceed the minimal number
of parametres on which the entanglement between any two sub-systems
depends. Shifting the LU optimisation from the final CM, back to the
engineering symplectic network, is in principle an extremely
involuted and nontrivial task.

This problem has been solved in Ref.~\cite{generic} for a special
subclass of Gaussian states, which is of null measure but still of
central importance for practical implementations. It is constituted
by those pure $n$-mode Gaussian states which can be locally put in a
standard form with null $\sig_{xp}$. This class encompasses
generalised GHZ-type Gaussian states, useful for CV quantum
teleportation networks \cite{network}, and Gaussian cluster states
\cite{zhang} employed in CV implementations of one-way quantum
computation \cite{menicucci}. It also comprises (as proven in the
previous Section) all three-mode pure Gaussian states \cite{3mpra},
whose usefulness for CV quantum communication purposes has been
thoroughly investigated \cite{3mj}. In the physics of many-body
systems, those states are quite ubiquitous as they are ground states
of harmonic Hamiltonians with spring-like interactions \cite{chain}.
For these Gaussian states, which we shall call here {\em
block-diagonal}, the minimal number of LU-invariant parametres
reduces to $n(n-1)/2$ for any $n$.\footnote{This number is easily
derived from the general framework developed in Sec.~\ref{condpuri}:
for $\sig_{xp}=0$, Eqs.~(\ref{first}) and (\ref{second}) reduce to
$\sig_{x}=\sig_{p}^{-1}$. The only further condition to impose after
the local reduction is then ${\rm diag}(\sig_{x})= {\rm diag}
(\sig_{x}^{-1})$, which brings the number of free parametres of the
symmetric $\sig_{x}$ from $(n+1)n/2$ down to $n(n-1)/2$}
Accordingly, one can show that an efficient scheme  can be devised
to produce block-diagonal pure Gaussian states, involving exactly
$n(n-1)/2$ optical elements which in this case are only constituted
by single-mode squeezers and beam-splitters, in a given sequence
\cite{generic}.

Borrowing the ideas leading to the state engineering of
block-diagonal pure Gaussian states, we propose here a scheme,
involving $(n^2-2n)$ independent optical elements, to produce more
general $n$-mode pure Gaussian states encoding correlations between
positions and momentum operators as well. To this aim, we introduce
`counter-beam splitter' transformations, named ``{\em
seraphiques}'', which, recovering the phase space ordering of
Sec.~\ref{techno}, act on two modes $j$ and $k$ as
$$
C_{j,k}(\vartheta) = \left(
\begin{array}{cccc}
 \cos (\vartheta ) & 0 & 0 & \sin (\vartheta ) \\
 0 & \cos (\vartheta ) & - \sin (\vartheta ) & 0 \\
 0 & \sin (\vartheta ) & \cos (\vartheta ) & 0 \\
 - \sin (\vartheta ) & 0 & 0 & \cos (\vartheta )
\end{array}
\right)\,.
$$
Such operations can be obtained by usual beam splitters (which we will denote by $B_{j,k}(\vartheta)$)
by applying a $\pi/2$ phase shifter $P_{k}$ on {\em only one} of the two considered modes.
$P_k$ is a local rotation mapping, in Heisenberg picture, $\hat{x}_{k}\mapsto-\hat{p}_{k}$
and $\hat{p}_{k}\mapsto\hat{x}_{k}$. In phase space, one has
$C_{j,k}(\vartheta)=P_{k}^{\sf T} B_{j,k}(\vartheta)P_{k}$.
Notice that, even though $C_{j,k}(\vartheta)$ is equal to the product of single-mode operations and beam
splitters, this does not mean that such a transformation is ``equivalent'' to a beam splitter in terms of
state generation. In fact, the local operations do not commute with the beam splitters, so that a product
of the kind $B_{j,k}(\vartheta')C_{j,k}(\vartheta'')$ {\em cannot} be written as $B_{j,k}(\vartheta)S_{l}$
for some local operation $S_{l}$ and $\vartheta$.

\begin{figure}[H]
\centering{\includegraphics[width=7cm]{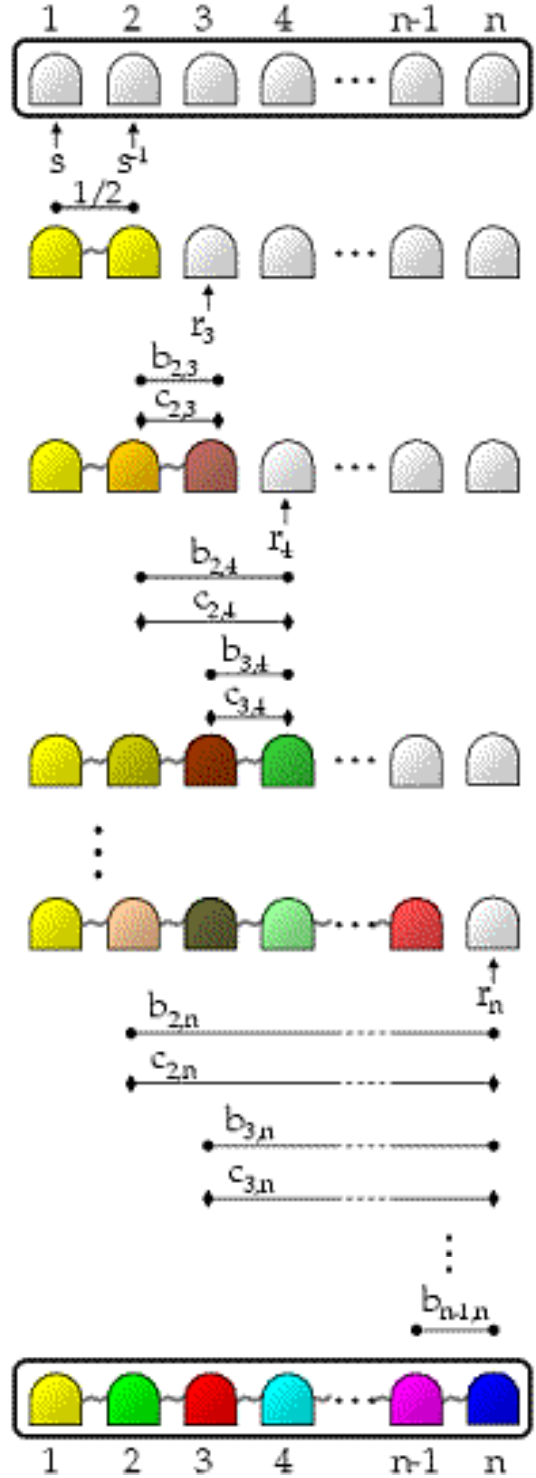}%
\caption{\label{schemino}(Color online) Possible scheme to create a
general $n$-mode pure Gaussian state. White shaded door-shaped ({\em
Schlutzkrapfen}-like) objects depict vacuum modes, while each color
corresponds to a different single-mode determinant ({\em
i.e.}~different degrees of local mixedness). Vertical arrows denote
single-mode squeezing operations with squeezing parametres $r_j$,
horizontal circle-ended lines denote beam-splitting operations
$b_{i,j}$ between modes $i$ and $j$, and horizontal diamond-ended
lines denote two-mode seraphiques parametrised by $c_{i,j}$. See
text for details.}}
\end{figure}

The state engineering scheme runs along the lines as the one for the
block-diagonal states, the only modification being that for each
pair of modes except the last one ($n-1,n$), a beam-splitter
transformation is followed by a seraphique. In more detail (see Fig.
\ref{schemino}): first of all (step i), one squeezes mode $1$ of an
amount $s$, and mode $2$ of an amount $1/s$ ({\em i.e.}~one squeezes
the first mode in one quadrature and the second, of the same amount,
in the orthogonal quadrature); then one lets the two modes interfere
at a $50:50$ beam splitter. One has so created a two-mode squeezed
state between modes $1$ and $2$, which corresponds to the Schmidt
form of the pure Gaussian state with respect to the $1 \times (n-1)$
bipartition. The second step basically corresponds to a
re-distribution, or allotment, of the initial two-mode entanglement
among all modes. This task can be obtained  by letting each
additional mode interact step-by-step with all the previous ones,
via beamsplitters and seraphiques (which are in turn combinations of
beam splitters and phase shifters). Starting with mode $3$ (which
was in the vacuum like all the subsequent ones), one thus squeezes
it (by an amount $r_3$) and combines it with mode $2$ via a
beam-splitter (characterized by a transmittivity $b_{2,3}$) and a
subsequent seraphique (parametrised by $c_{2,3}$). Then one squeezes
mode $4$ by $r_4$ and lets it interfere sequentially both with mode
$2$ (via a beamsplitter with $b_{2,4}$ and a seraphique with
$c_{2,4}$) and with mode $3$ ($b_{3,4}$ and $c_{3,4}$). This process
can be iterated for each other mode, as shown in
Fig.~\ref{schemino}, until the last mode $n$ is squeezed ($r_n$) and
entangled with the previous ones via beam-splitters with respective
transmittivities $b_{i,n}$, $i=2,\cdots,n-1$, and corresponding
seraphiques with amplitudes $c_{i,n}$, $i=2,\cdots,n-2$. We remark
that mode $1$ becomes entangled with all the other modes as well,
even if it never comes to a direct interaction with each of modes
$3,\cdots,n$.

%%%%%%%%%%%%%%%%%%%%%%%%%%%%%%%%%%%%%%%%%%%%%%%%%%%%%%%%%%%

This scheme is implemented with minimal resources. Namely, the state
engineering process is characterised by one squeezing degree (step
i), plus $n-2$ individual squeezings, together with
$\sum_{i=1}^{n-2} i = (n-1)(n-2)/2$ beam-splitter transmittivities,
and $[\sum_{i=1}^{n-2} i]-1 = n(n-3)/2$ seraphique transmittivities,
which amount to a total of $(n^2-2n)$ quantities, exactly the ones
parametrising a general pure Gaussian state of $n\ge3$ modes up to
local symplectic operations. While this scheme is surely more
general than the one for block-diagonal states, as it enables to
efficiently create a broader class of pure Gaussian states for
$n>3$, we shall leave it as an open question to check if the recipe
of Fig.~\ref{schemino} is general enough to produce {\em all} pure
$n$-mode Gaussian states up to LUs. Verifying this analytically
leads to pretty cumbersome expressions already for $n=4$. Instead,
it would be very interesting to investigate if the average
entanglement of the output Gaussian states numerically obtained by a
statistically significant sample of applications of our scheme with
random parametres, matches the {\em typical} entanglement of pure
Gaussian states under ``thermodynamical'' state-space measures as
computable along the lines of Ref.~\cite{typical}. This would prove
the optimality and generality of our scheme in an operational way,
which is indeed more useful for practical applications.

\section{Epilogue}

In view of the previous, comprehensive characterisation of
structural and informational properties of pure $n$-mode Gaussian
states under LU operations, it is natural to question if the
$n(n-3)/2$ additional parametres encoded in $\hat x$-$\hat p$
correlations for non-block-diagonal states, have a definite impact
on the bipartite and multipartite entanglement.

\begin{figure}[t!]
\centering{\includegraphics[width=10cm]{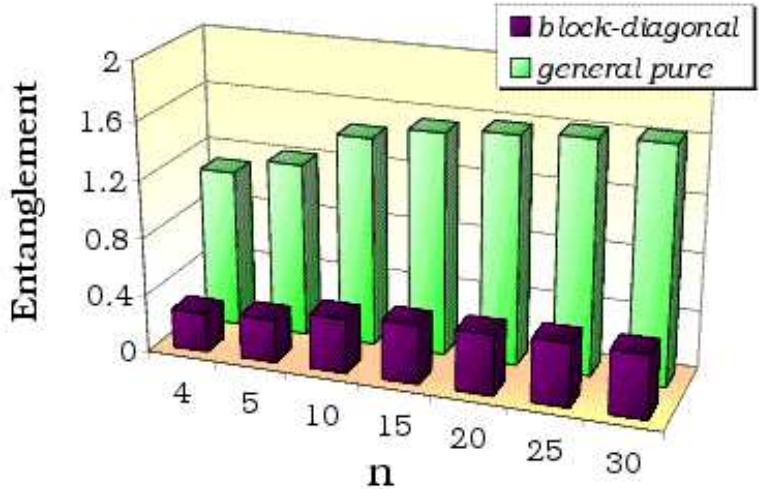}%
\caption{\label{istotipo}(Color online) Typical entanglement,
measured by the Von Neumann entropy, between one mode and the
remaining $n-1$ modes in two classes of pure $n$-mode Gaussian
states, for $n \ge 4$. Light green bars denote completely general
pure states, while dark purple bars refer to block-diagonal pure
states. For each $n$, the entanglement is averaged over 10000 random
realisations of pure Gaussian states (with and without direct $\hat
x$-$\hat p$ correlations, respectively) according to the
microcanonical state space measure introduced in \cite{typical}, at
a fixed total energy ${\cal E}=5n$. Nonvanishing correlations
between position and momentum operators in the covariance matrix,
clearly yield an increase in the typical entanglement of pure
Gaussian states, more evident with increasing number $n$ of modes.}}
\end{figure}

 At present,
usual CV protocols are devised, even in multimode settings, to make
use of states without any $\hat x$-$\hat p$ correlations. In such
cases, the economical (relying on $(n-1)n/2$ parametres)
``block-diagonal state engineering'' scheme detailed in
Ref.~\cite{generic} is clearly the optimal general strategy for the
production of entangled resources. However, theoretical
considerations strongly suggest that states with $\sig_{xp}\neq0$
might have remarkable potential for improved quantum-informational
applications. In fact, considering again the thermodynamical
entanglement framework of  Gaussian states \cite{typical}, one can
define natural averages either on the whole set of pure Gaussian
states, or restricting to states with $\sig_{xp}=0$.\footnote{The average over the whole set of
pure Gaussian states is realised by integrating over the Haar measure of the compact subgroup $K(n)$,
isomorphic to $U(n)$. The restriction to all the states with vanishing $xp$ block is instead
achieved by considering only orthogonal symplectic transformations of the form $R\oplus R$
with $R\in O(n)$ -- which form a group isomorphic to $O(n)$ -- and by integrating over the Haar measure of $O(n)$ [as opposed to $U(n)$].}
Well, numerics
unambiguously show (see Fig. \ref{istotipo}) that the average
entanglement (under any bipartition) of Gaussian states without
$\hat x$-$\hat p$ correlations (like the ones considered in
\cite{generic}) is systematically lower than the typical
entanglement of more general pure Gaussian states, with this
behaviour getting more and more manifest as the total number of
modes increases (clearly, according to what we have shown in this
work, this only occurs for $n>3$). In a way, the full entanglement
potential of Gaussian states is diminished by the restriction to
block-diagonal states.

%%%%%%%%%%%%%%%%%%%%%%%%%%%%%%%%%%%%%%%%%%%%%%%%%%%%%%%%%%%

On the other hand,  the comparison between the average entanglement
generated in randomising processes based on the engineering scheme
proposed here and the block diagonal one is under current
investigation as well. If the present scheme turned out to be
out-performing the previous ones in terms of entanglement generation
-- as expected in view of the argument above -- this would be a spur
to the exploration of novel CV protocols,
capable of adequately exploiting $\hat x$-$\hat p$ correlated
resources.

\bigskip
\noindent{\bf Acknowledgments}\\
We thank N.~Schuch and M.~M.~Wolf for pointing out an independent
derivation \cite{gaussianeof} of the counting argument employed in
Section~\ref{condpuri}. We are grateful to M.~B.~Plenio,
F.~Illuminati, and C.~Dawson for helpful discussions. The pictorial
and terminological consultancy of D.~Gross has been invaluable (see
Figure \ref{schemino}).

%%%%%%%%%%%%%%%%%%%%%%%%%%%%%%%%%%%%%%%%%%%%%%%%%%%%%%%%%%%

\end{document}